\RequirePackage[hyphens]{url}
\documentclass[12pt]{article}
\usepackage{pdflscape}
\usepackage{tikz,natbib,xfrac,amsmath,amsthm,booktabs,subfigure,anysize,hyperref,graphicx,float,tabularx,multirow}
\usepackage[margin=1in]{geometry}
\usepackage{xcolor}
\usepackage{bm}
\definecolor{dark-red}{rgb}{0.4,0.15,0.15}
\definecolor{dark-blue}{rgb}{0.15,0.15,0.4}
\definecolor{medium-blue}{rgb}{0,0,0.5}
\definecolor{ChadBlue}{rgb}{.1,.1,.5}
\definecolor{ChadDarkBlue}{rgb}{.1,0,.2}
\definecolor{ChadBlue}{rgb}{.1,.1,.5}
\definecolor{ChadRoyal}{rgb}{.2,.2,.8}
\definecolor{ChadGreen}{rgb}{0,.4,0}    
\definecolor{ChadRed}{rgb}{.5,0,.5}  
\hypersetup{
    colorlinks, linkcolor={ChadRed},
    citecolor={ChadBlue}, urlcolor={medium-blue}
}

\usepackage[font=small,labelfont=bf,justification=justified,singlelinecheck=false]{caption}
\usepackage{amssymb}
\usepackage{amsfonts}
\usepackage{amsmath}
\usepackage{setspace}  
\usepackage[all]{hypcap}        

\graphicspath{{figures/}}
\usepackage[english]{babel}
\usepackage{longtable}
\usepackage[singlelinecheck=off]{caption}
\usepackage[singlelinecheck=false]{caption}
\usepackage{array}
\usepackage{float}

\usepackage [autostyle, english = american]{csquotes}
\MakeOuterQuote{"}
\usepackage{chngcntr}
\usepackage{rotating, graphicx}
\usepackage{epstopdf}
\usepackage{threeparttable}
\usepackage{sectsty}
\sectionfont{\large}

\usepackage{longtable}
\usepackage{booktabs}
\usepackage{array}
\newcolumntype{R}{>{\raggedleft\arraybackslash}p{2.5cm}}
\newcolumntype{Q}{>{\raggedright\arraybackslash}p{14cm}}
\newcolumntype{U}{>{\raggedright\arraybackslash}p{11cm}}

\usepackage{tabu}
\usepackage{dcolumn}
\usepackage{placeins}
\usepackage[T1]{fontenc}

\title{Impact of COVID-19 on the trade of goods and services in Spain\thanks{I thank Marina Abad, Federico Carril-Caccia, Juan de Lucio, Carmen Díaz-Mora, Benedikt Heid, Bernardo Hern\'{a}ndez, Bart Kamp, César Martín, Ra\'{u}l M\'{i}nguez, Rafael Myro, Jon Ortueta, Francisco Requena, Mikel Trojaola, Albert Vinaixa, an anonymous reviewer, and participants at II Encuentro Red de Investigadores en Internacionalizaci\'{o}n Banco de Espa\~na for their very valuable comments and suggestions. I also thank Barth\'{e}l\'{e}my Bonadio and Nitya Pandalai-Nayar for providing me with the estimations for Spain in \cite{bonadio2020covid}. I gratefully acknowledge the financial support from the Spanish Ministry of Science, Innovation and Universities (RTI2018-100899-B-I00, co-financed by FEDER) and the Basque Government Department of Education, Language policy, and Culture (IT885-16).}}

\author{\large {Asier Minondo}\thanks{Deusto Business School, University of Deusto, Camino de Mundaiz 50, 20012 Donostia - San Sebasti\'{a}n (Spain). Email: \href{mailto:aminondo@deusto.es}{aminondo@deusto.es}}}

\date{ \today \\  }

\usepackage[normalem]{ulem}

\begin{document}

\maketitle

\begin{abstract}
	The COVID-19 crisis has led to the sharpest collapse in the Spanish trade of goods and services in recent decades. The containment measures adopted to arrest the spread of the virus have caused an especially intense fall of trade in services. Spain's export specialization in transport equipment, capital and outdoor goods, and services that rely on the movement of people has made the COVID-19 trade crisis more intense in Spain than in the rest of the European Union. However, the nature of the collapse suggests that trade in goods can recover swiftly when the health crisis ends. On the other hand, COVID-19 may have a long-term negative impact on the trade of services that rely on the movement of people.
\end{abstract}

\begin{flushleft}
\end{flushleft}

\newpage 
\doublespacing

\section{Introduction}
\label{sec:introduction}

There is little doubt that 2020 will be remembered as the COVID-19 year. The virus that emerged in China has, thus far, infected more than 51 million and killed almost 1.3 million people globally.\footnote{Johns Hopkins Coronavirus Resource Center. Available at: \url{https://coronavirus.jhu.edu/map.html}. Accessed November 11, 2020.} To avert the spread of the virus, governments have introduced strict social-distancing and containment measures, and these have had a large negative impact on global economic activity. The International Monetary Fund (IMF) projected a 4.4\% drop in the global gross domestic product (GDP) in 2020, the largest decrease in 40 years \citep{imf2020weoctober}. This reduction in global economic activity has led to a vast decrease in the global flow of goods and services. The World Trade Organization reported a 14.3\% decline in the global volume trade of goods for the second quarter of 2020 compared with the first quarter of 2020. This decrease was larger than the one provoked by the Great Recession between the third quarter of 2008 and first quarter of 2009 (10.2\% decline). 

This paper analyzes the impact of COVID-19 on the trade of goods and services in Spain. The pandemic has already led to the steepest decline in trade in recent decades, reducing the value of Spanish trade flows to 2016 figures. Furthermore, owing to the large share of transport equipment, capital goods, products that are consumed outdoors (i.e., outdoor goods), and tourism in Spanish exports, and the fact that demand for these products and services decreased notably during the COVID-19 crisis, the pandemic has had a larger negative effect in Spain than in the rest of the European Union (EU).

The paper begins with a chronology of the trade collapse generated by COVID-19 in Spain. Although the first news about the virus emerged in January 2020, the trade crisis did not begin until confinement measures were introduced in Spain and its main trading partners in March. The most severe phase of the crisis occurred between the second half of March and May 2020, when very strict containment measures were implemented. However, the year-to-year decreases in trade flows began to soften from June 2020 onward. On the other hand, as this article goes to press, the latest available monthly trade figures of August were still lower than those at the same point in 2019. 

Second, to understand the severity of the trade collapse, I compare the drop in exports during the COVID-19 crisis with the aforementioned Great Recession of 2008-2009. I show that exports dropped faster during the COVID-19 crisis, but that the total drop was lower during the COVID-19 crisis. Regarding services, the first three months of the COVID-19 crisis were enough to reach the same overall collapse. In both cases, Spanish firms continued to sell to their foreign customers, but at lower amounts. Because Spanish firms maintained their relationship with foreign customers, it is reasonable to expect a rapid recovery of good exports after the health crisis ends.

Third, I show that COVID-19 had a larger negative impact on Spain's exports than on those of the rest of the EU. I further find that the country distribution of Spanish exports contributed to this differential impact. I also illustrate that COVID-19 had an heterogeneous impact on Spanish regions' export of goods.

Fourth, I analyze whether Spanish firms' dependence on imported goods contributed to the COVID-19 trade crisis. I show that a large share of Spanish imports was sourced by importers that only had a single supplier country. Despite this dependence, Spanish exporters did not identify disruptions in the supply chain as a major factor that drove the decrease in trade flows. However, the large increase in demand for medical materials and equipment, and the fact that Spain already depended on imports before the COVID-19 crisis led to severe sourcing problems for these products during the first stages of the health crisis.

Fifth, I speculate whether the COVID-19 crisis will cause permanent changes to Spanish trade. I argue that it may lead to changes in the composition of service exports, have a negative impact on the growth of Spanish exports in the medium term, and increase the concentration of Spanish exports in EU countries. The last section concludes this paper.

\section{Chronology of a trade crisis}
\label{sec:chronology}

The first significant negative impact of COVID-19 on Spanish trade occurred on February 12, 2020, when the organizers of the World Mobile Congress, a major event contributing to service exports, announced its cancellation because of health risks. That month, firms began to worry that a lack of intermediate inputs manufactured in China would lead to production shutdowns \citep{expansion2020covid}. These concerns further increased when some northern Italian regions that hosted factories that exported components to Spain were put under quarantine at the end of February \citep{vozpopuli2020covid}. However, it was in March, in response to the declaration of the state of alarm in Spain and the introduction of confinement measures in many other countries, when COVID-19 began to have a significantly negative impact on trade flows.

\begin{figure}[p!]
	\begin{center}
		\caption{Impact of COVID-19 on Spanish trade  (Millions of euros; current values)}
		\label{fig:chronology}
		\hspace*{-1cm}\begin{tabular}{cc}  
			\multicolumn{2}{c}{Panel A: Goods}  \\
			A1. Exports & A2. Imports \\
			\includegraphics[height=2.5in]{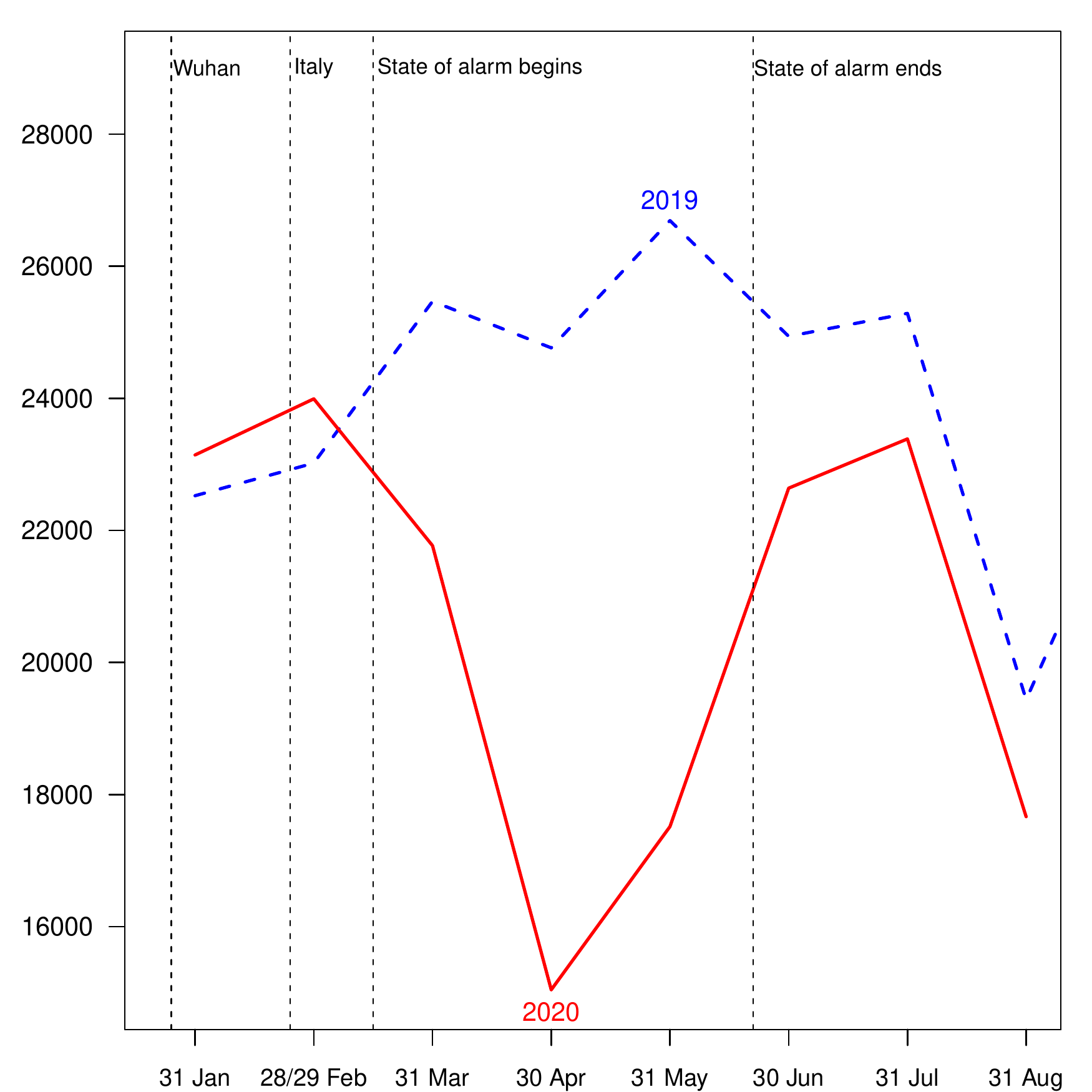} & 
			\includegraphics[height=2.5in]{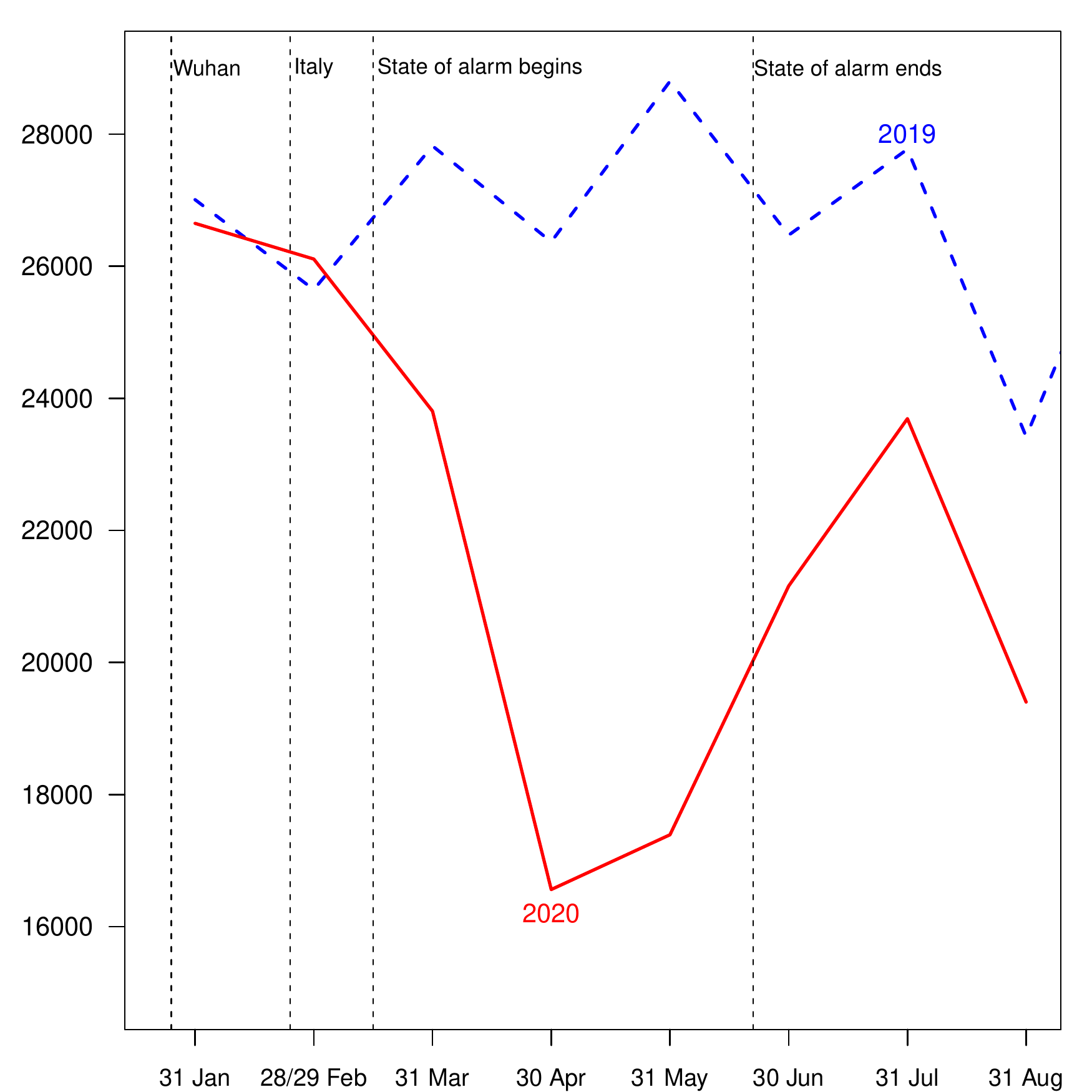} \\
			\multicolumn{2}{c}{Panel B: All services, including tourism}  \\
			B1. Exports & B2. Imports \\
			\includegraphics[height=2.5in]{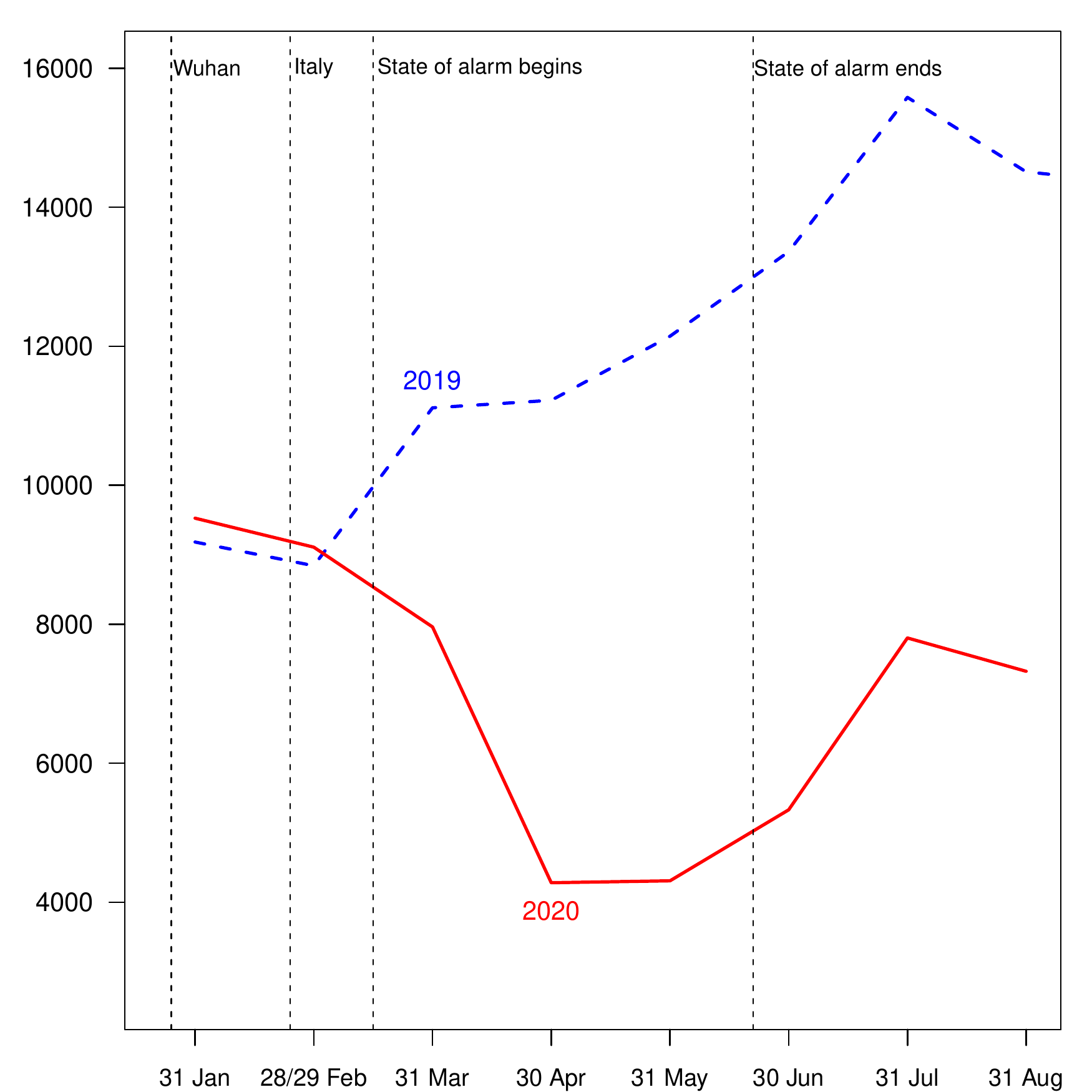} & 
			\includegraphics[height=2.5in]{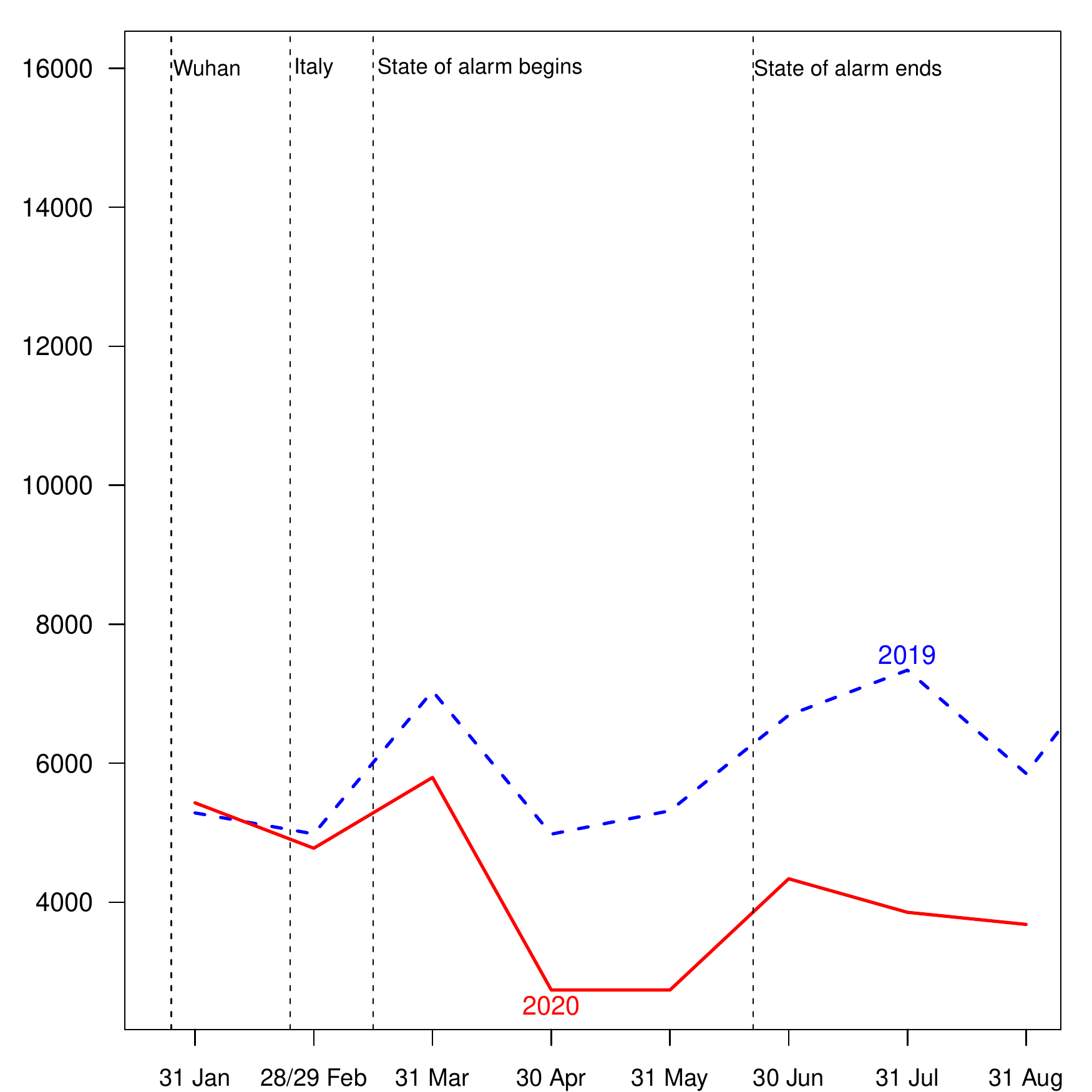} \\
			\multicolumn{2}{c}{Panel C: Tourism}  \\
			C1. Exports & C2. Imports \\
\includegraphics[height=2.5in]{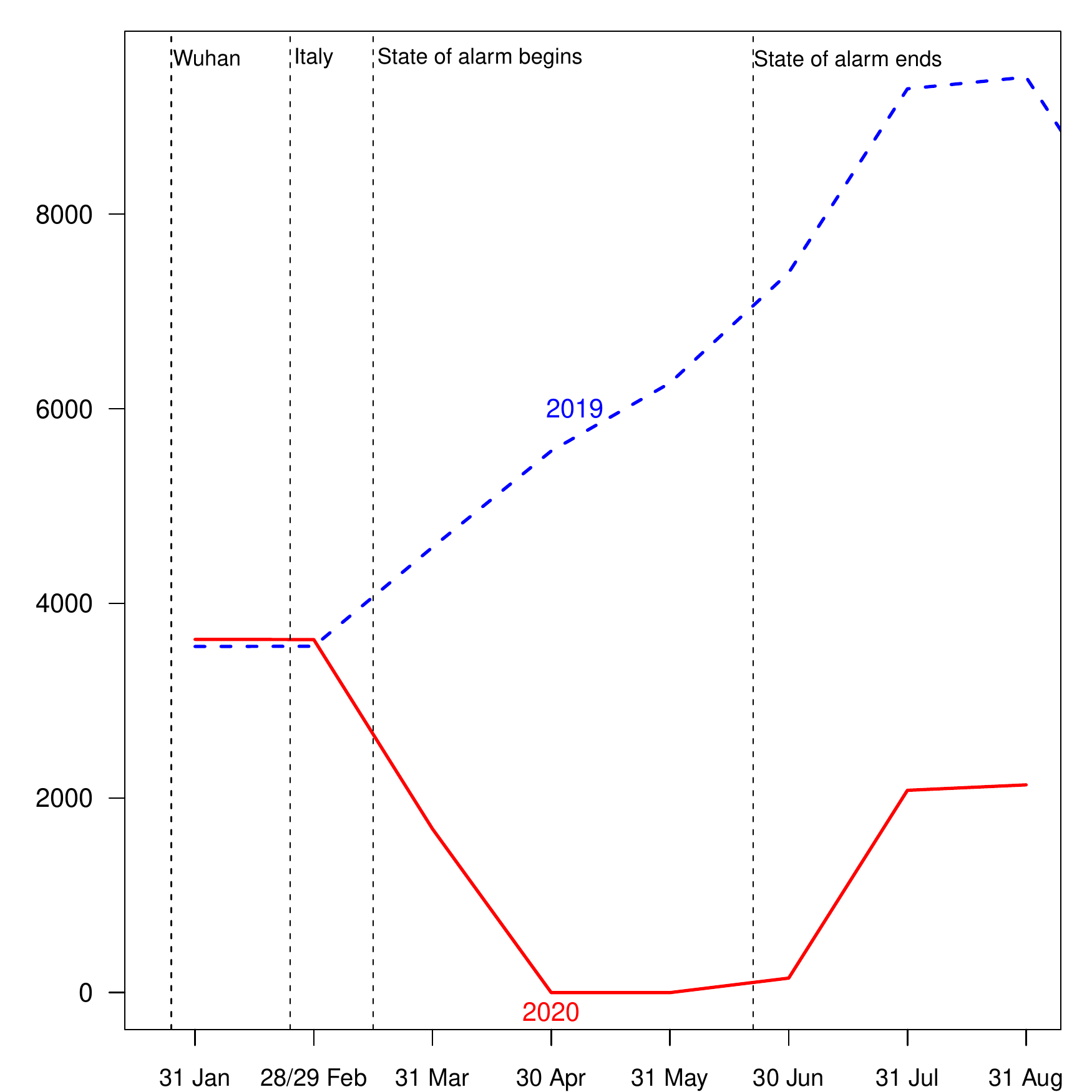} & 
\includegraphics[height=2.5in]{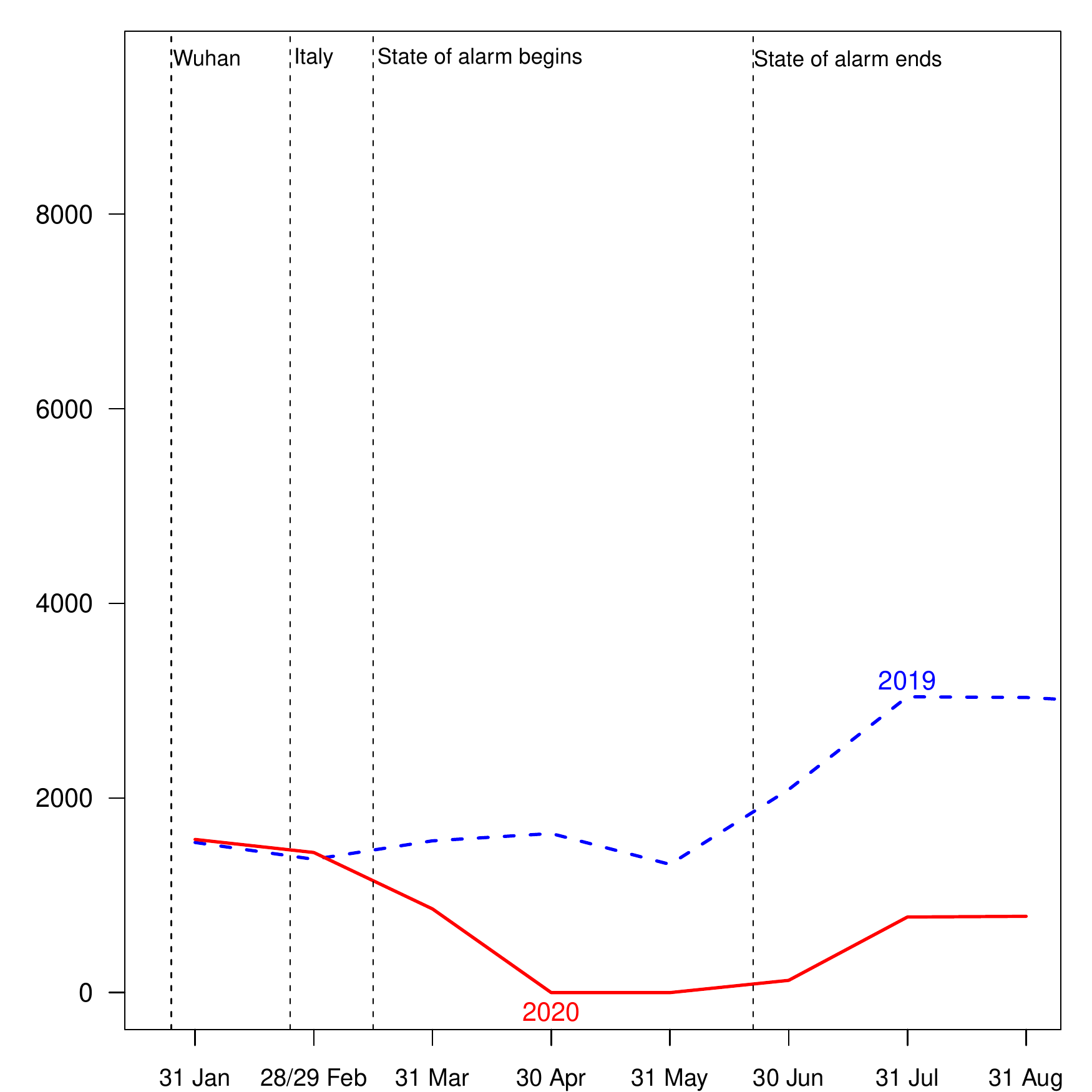} \\

		\end{tabular}
	\end{center}
	\footnotesize{Source: Author's own elaboration based on Customs and Bank of Spain data.}
\end{figure}

Panel A1 of Figure~\ref{fig:chronology} shows the evolution of good exports in 2019 and 2020.\footnote{Trade data for goods was obtained from the Customs and Excise Department of the Spanish Tax Agency's database. It can be downloaded at: \url{https://www.agenciatributaria.es/AEAT.internet/Inicio/La_Agencia_Tributaria/Memorias_y_estadisticas_tributarias/Estadisticas/_Comercio_exterior_/Datos_estadisticos/Descarga_de_Datos_Estadisticos/Descarga_de_Datos_Estadisticos.shtml}. Data can also be accessed via queries from the State Secretary of Commerce DataComex interface at: \url{http://datacomex.comercio.es/principal_comex_es.aspx}. Figures correspond to the trade declared by firms each month. Note that these figures are provisional, because a month's figure may include shipments that occurred at previous months and were not declared when they happened. Monthly figures are revised and become definitive with an approximately two-year lag.} Goods represented two-thirds of all Spanish export values in 2019. The blue dash and the red solid lines are monthly exports in 2019 and 2020, respectively. Exports in January and February were 2.7\% and 4.2\% higher than during the same months of 2019. With the introduction of containment measures in many EU countries (the most important destinations of Spanish exports), the value of exports decreased by 14.5\%, 39.3\%, and 34.3\% in March, April, and May, respectively, compared to the previous year. Coinciding with the relaxation of containment measures at the destinations of Spanish exports, the decreases began to smoothen: exports ``only'' dropped by 9.2\%, 7.5\%, and 9.1\% in June, July, and August, respectively.\footnote{\cite{delucio2020covidcollapse}, using econometric analyses, confirm the correlation between the drop in the value of exports and the stringency of the containment measures adopted by Spanish trading partners.}

COVID-19 also manifested as a shock in the import of goods (Panel A2). In January 2020, imports were 1.3\% lower than during the same month of 2019. However, in February, even when lockdown measures had already been adopted in China and some northern Italian regions, imports were 1.8\% greater than during the same month of 2019. With the introduction of confinement measures in Spain, imports dropped by 14.4\% in March, 37.2\% in April, and 39.6\% in May. Spain's state of alarm ended on June 21, 2020, and the year-to-year drop in imports was lower than that in the previous two months. Nevertheless, it was still very large: 20.1\%. Imports increased in July relative to June. However, they remained much lower than during the same period of 2019 (-14.7\%). In August, exports remained 17.2\% lower than during the same month of 2019. During the March-August period, import of goods decreased more than exports. This can be explained by the larger drop in economic activity in Spain than that of its trading partners.\footnote{According to INE and Eurostat data, during the second quarter of 2020, Spanish GDP decreased by 17.8\% relative to the first quarter of 2020. In the EU, the most important Spanish trading partner, the drop was of 11.5\%.}

The negative impact of COVID-19 on trade was greater for services than for goods (Panels~B1 and B2 of Figure~\ref{fig:chronology}).\footnote{Data on service exports and imports were obtained from the Bank of Spain, Balance of Payments and International Investment Position database, available at \url{https://www.bde.es/webbde/en/estadis/infoest/temas/sb_extbppii.html}.} Many services require the movement of the supplier or the customer, and the confinement measures made such movement across national boundaries almost impossible. Export of services decreased by more than 60\% between April and June, and almost by half in July and August relative to the same months of 2019. However, the drop in imports, although large, was smaller than in exports. 

Tourism, which represented 49.4\% and 32.7\% of service exports and imports in 2019, respectively, was the main contributor to the massive reduction in the trade of services (Panels~C1 and C2 of Figure~\ref{fig:chronology}). In April and May 2020, tourism receipts and payments were zero. In June, when hotels were allowed to reopen in Spain, and border controls were lifted in the rest of Europe, tourism receipts and payments were still 98.0\% and 94.1\% lower, respectively, than during the same period of 2019. Although tourism receipts and payments recovered in July and August, they remained around three-quarters lower than those of the same period of 2019. 

The decrease in non-tourism services trade, comprising transport, finance, communications, and business services, was much lower than that of tourism. Between March and August 2020 exports and imports of non-tourism services decreased by 12.6\% and 16.6\%, respectively. Figures for tourism export and imports were 85.8\% and 79.9\%, respectively. The Survey of International Trade in Services provided by the National Statistics Institute reported an increase in business and financial service exports during the second quarter of 2020 relative to the same period of 2019. Contrarily, there was a sharp decrease in transport-related exports. Communications and insurances were the only branches showing positive import growth during the second quarter of 2020.\footnote{The survey can be accessed at \url{https://ine.es/dynt3/inebase/index.htm?padre=2124&capsel=2124}}

Overall, during the March-August period of 2020, exports and imports of goods decreased by 19.5\% and 24.1\%, respectively, relative to the same period of 2019. Service exports and imports decreased by 52.5\% and 37.8\%, respectively. If we add exports to imports and merchandises to services, total Spanish trade decreased by 28.9\%. This was the largest biannual drop in Spanish trade in decades. 

\begin{figure}[t]
	\begin{center}
		\caption{Numbers of exporters and importers of goods}
		\label{fig:traders}
		\hspace*{-1cm}\begin{tabular}{cc}
			(A) Exporters & (B)  Importers \\
			\includegraphics[height=3.0in]{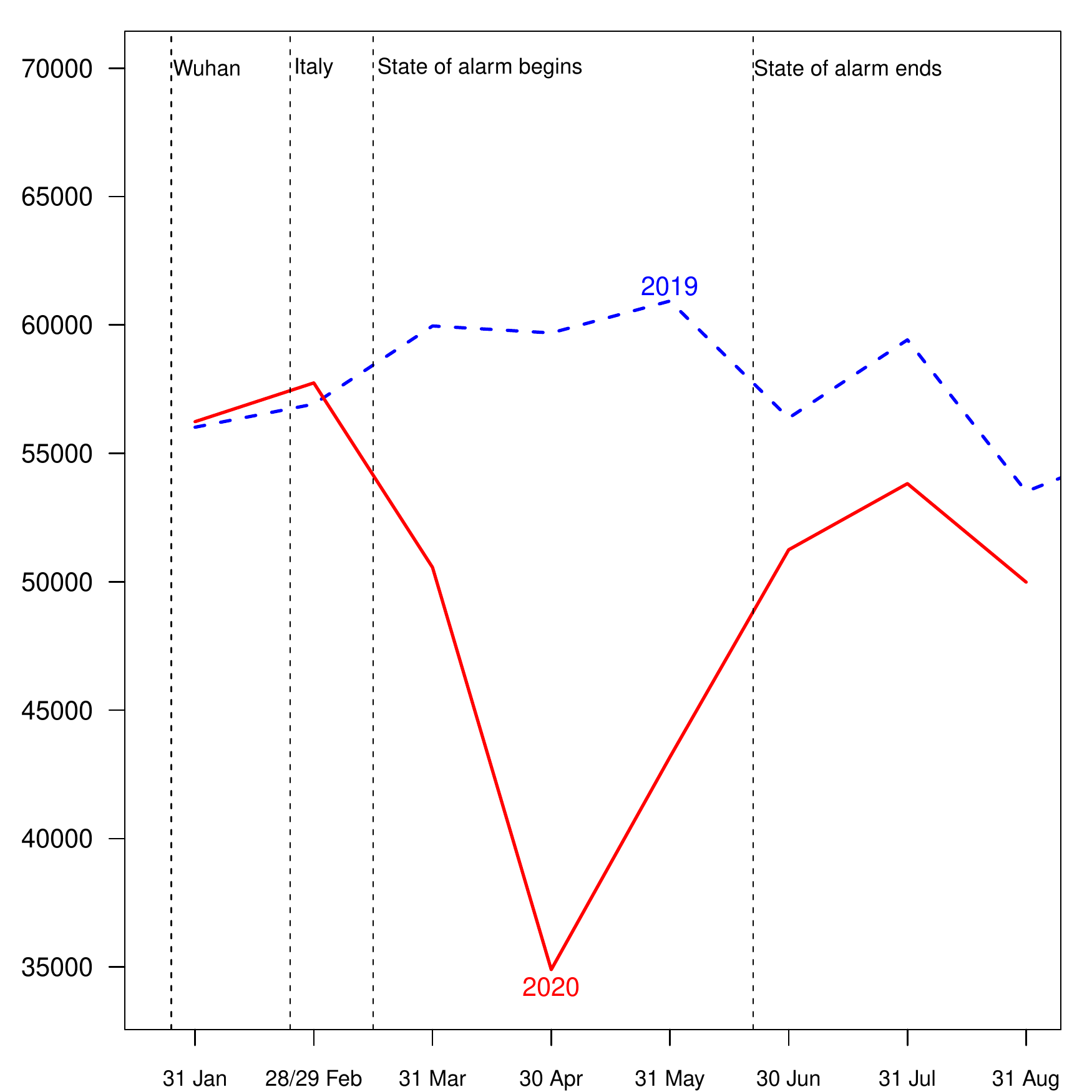} & 
			\includegraphics[height=3.0in]{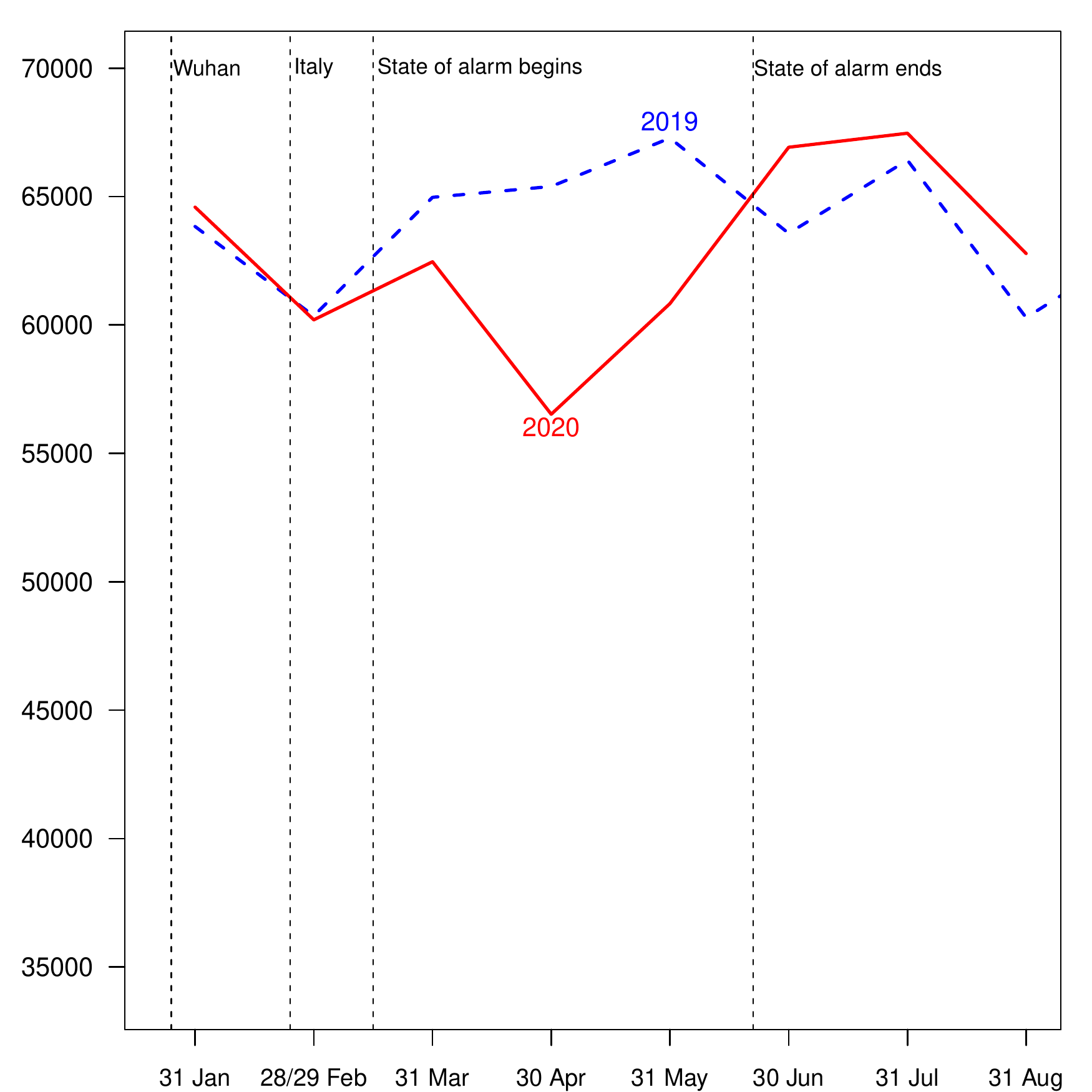} 
		\end{tabular}
	\end{center}
	\footnotesize{Source: Author's own elaboration based on data from Datacomex.}
\end{figure}

COVID-19 also had a negative impact on the number of firms exporting and importing goods (Figure~\ref{fig:traders}).\footnote{\label{foot:firm_definition}The Department of Customs and Special Taxes of the Revenue Agency defines a firm as an operator that conducts declared international trade operations. An operator might not have the legal status of a firm.} During March and April, there was a decrease in the number of traders, especially exporters. However, from May onward, there was a very fast recovery in that number. In August 2020, the number of exporters was 6.6\% lower than that during the same month of 2019. Contrarily, the number of importers was already higher than that of 2019. 

The drop in the number of exporters would be more worrisome if it had affected regular exporters, which accounted for 95.1\% of Spanish exports in 2019.\footnote{Regular exporters are defined as firms that exported without interruption in the last four years.} However, \cite{mineco2020informeagosto} showed that during the period of January-August 2020, the number of regular exporters (51,370) was 1\% higher than that during the same period of 2019.\footnote{There were many regular exporters that exported a very small amount per year. I recalculated the number of regular exporters removing all export transactions with a value lower than 1,500 euros. The number of regular exporters decreased from 37,873 in January to 37,519 in August, which represented a 1\% fall.} These figures indicate that the drop in the number of exporters could be explained by the exit of firms that only occasionally exported .

\section{COVID-19 vs. the Great Recession}
\label{sec:versus}

To better evaluate the extent and nature of the impact of COVID-19 on Spanish exports, I compare it with the Great Recession of 2008-2009. To identify the beginning of a crisis, for each year and month, I compute a moving window totaling the value of exports in the actual month and the previous eleven months:

\begin{equation}
	\label{eq:window}
	Window_{m,t}=\sum_{m-11}^{m}X_{m,t'}
\end{equation}

where $X_{m,t'}$ is the value of exports in the month $m$ in year $t'$, and ${t'~\in(t-1,t)}$. I define the beginning of a crisis when the moving window is lower than that of the previous month, but is higher than that of the following month ($Window_{m-1,t}>Window_{m,t}>Window_{m+1,t}$). I give the value of 100 to $Window_{m,t}$ when $m$ is the month prior to the beginning of the crisis. Using this definition, the Great Recession began in October 2008 for goods and one month later for services. The COVID-19 crisis began in March 2020 for both goods and services.

\begin{figure}[htbp]
	\begin{center}
		\caption{COVID-19 vs. the Great Recession (Exports=100 in the month previous to the start of the crisis)}
		\label{fig:covidvsgreat}
		\hspace*{-1cm}\begin{tabular}{cc}
			(A) Goods & (B)  Services \\
			\includegraphics[height=3.0in]{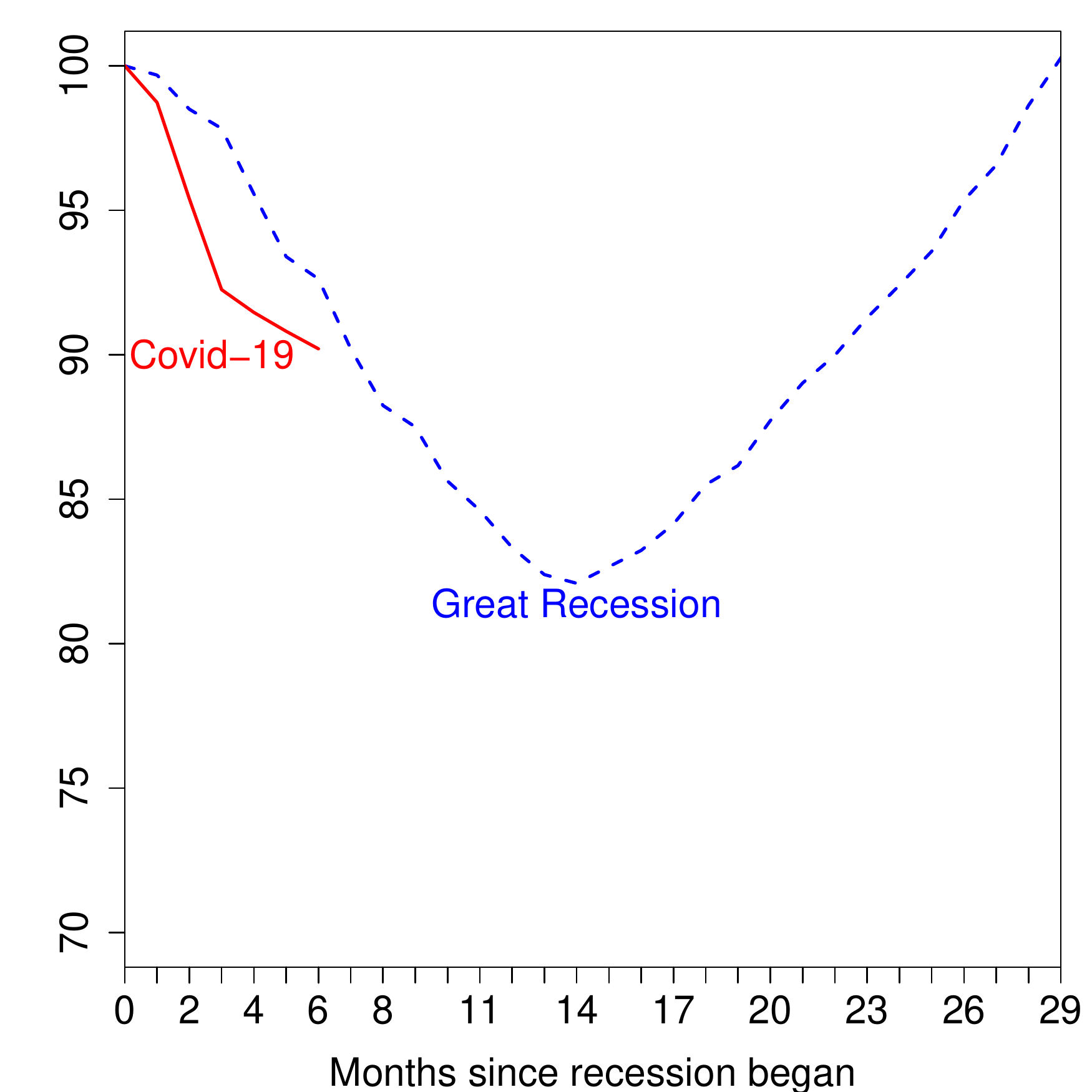} & 
			\includegraphics[height=3.0in]{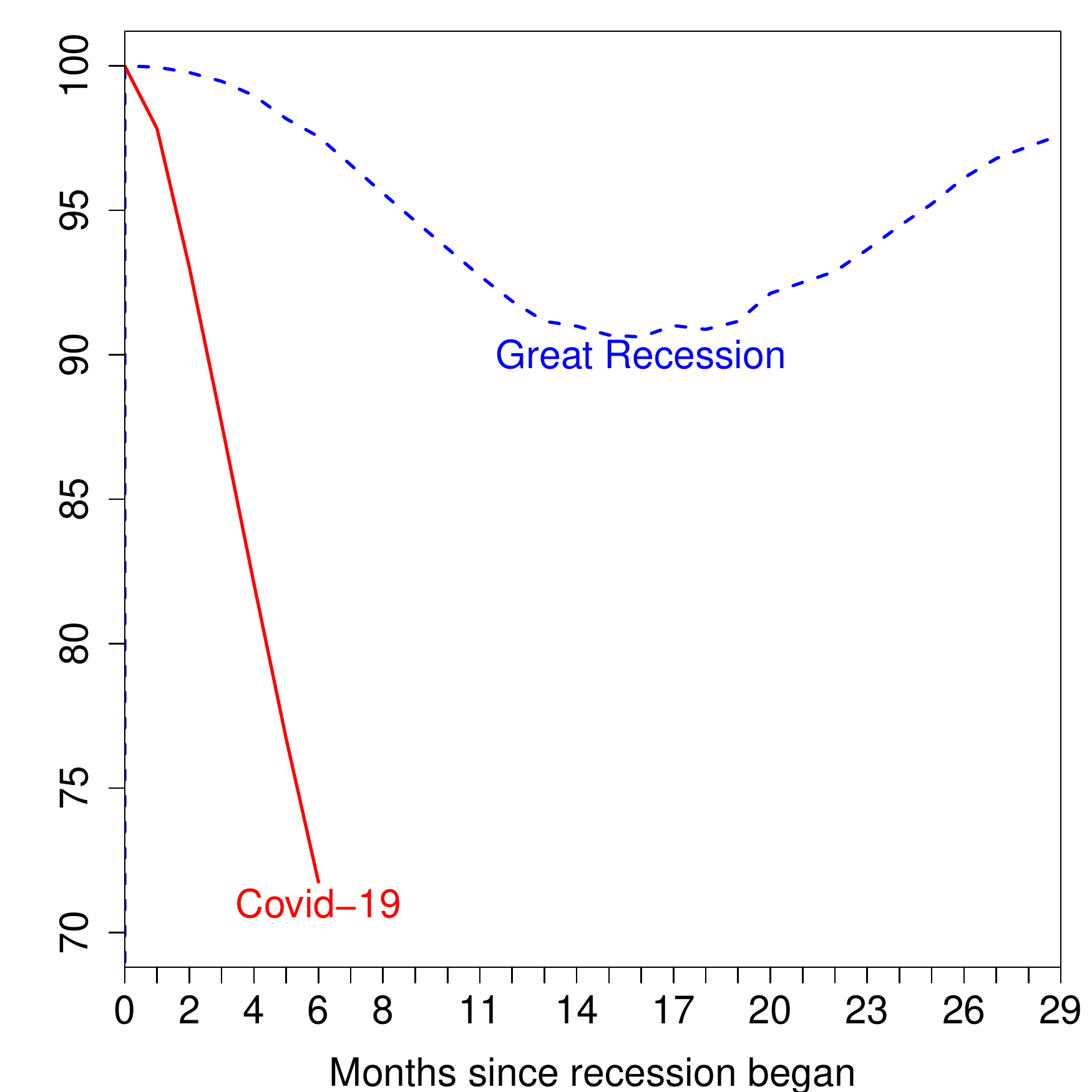} 
		\end{tabular}
	\end{center}
	\footnotesize{Source: Author's own elaboration based on data from Customs and Bank of Spain.}
\end{figure}

Panel~(A) of Figure~\ref{fig:covidvsgreat} shows that the export-of-goods index decreased faster during the COVID-19 crisis than during the Great Recession. However, until August 2020, the total decrease in the COVID-19 crisis was smaller than that of the Great Recession. Regarding services (panel~(B) of Figure~\ref{fig:covidvsgreat}), the decrease in the exports index during the COVID-19 crisis was also faster than that of the Great Recession. The intensity of the COVID-19 trade crisis was so severe for services that after the first three months, March-May 2020, the drop in the exports index was already larger than that of the entire Great Recession. Furthermore, the drop in the index of service exports during the COVID-19 crisis overcame the drop in the export-of-goods index during the Great Recession.

I also compare the sources of the trade collapse in the Great Recession with those of the COVID-19 crisis. The decrease in the value of exports and imports can be decomposed in two margins: intensive and extensive. The intensive margin is the reduction in exports when a firm continues to export a product to a destination and sales are reduced. The extensive margin comprises two sub-margins. Sub-margin (a), denoted by extensive - products\&destinations, is the reduction in exports when a firm ceases to sell a product in a destination. Sub-margin (b), denoted by extensive - firms, is the reduction in exports when a firm ceases to export. The definition of margins is equivalent for imports.

Recovery from a crisis can be faster if the intensive margin is the main contributor to the decrease in exports and imports. In this margin, the relationship between the exporter (impoter) and its foreign customer (supplier) remains alive, and exports (imports) recover quickly when the slump ends. Contrarily, if the adjustment comes from the extensive margin, many trade relationships will be lost. Firms will then need to invest in recovery efforts and create new export (import) relationships. Because these investments are costly, some firms cannot return to their foreign markets, even when demand recovers.\footnote{\label{foot:references}\cite{roberts1997decision} showed that a firm has a higher probability to serve a destination if it did it the previous year. \cite{defever2015spatial} concluded that this effect declines rapidly and vanishes after about two years.} For example, \cite{behrens2013tradecrisis} concluded that the swift recovery in exports in Belgium after the Great Recession occurred because 97\% of the collapse was accounted for the intensive margin. For Spain, \cite{delucio2011crisis} calculated that 90\% of the trade collapse during the Great Recession occurred in the intensive margin, which, as anticipated, led to a swift recovery of Spanish exports.

\begin{figure}[h!]
	\begin{center}
		\caption{Contribution of the intensive and extensive margins to the decrease in the trade of goods: COVID-19 vs. Great Recession}
		\label{fig:intensive_extensive}
		\begin{tabular}{c c}
		Panel A. Exports & Panel B. Imports\\
		\includegraphics[height=3.0in]{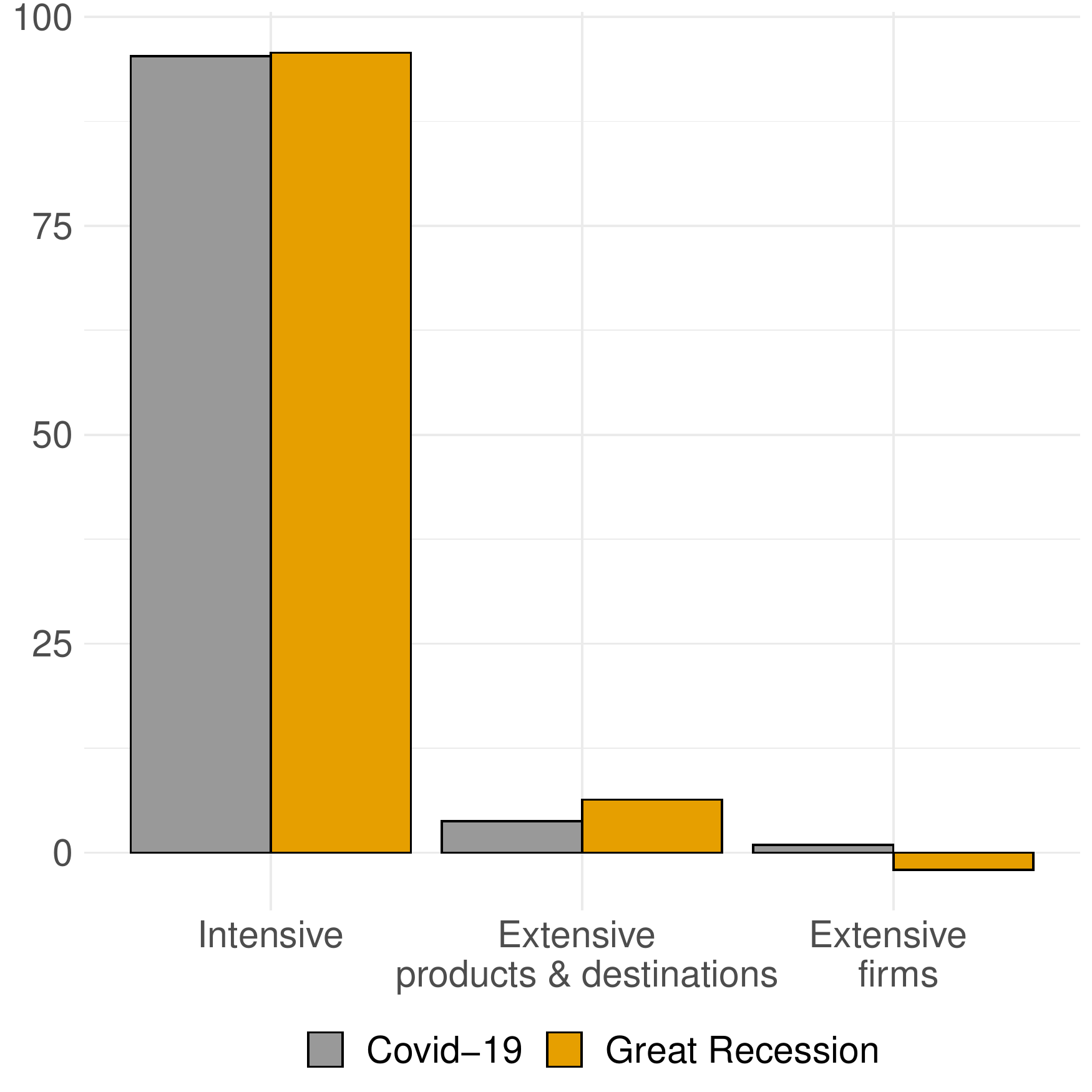}	& 		\includegraphics[height=3.0in]{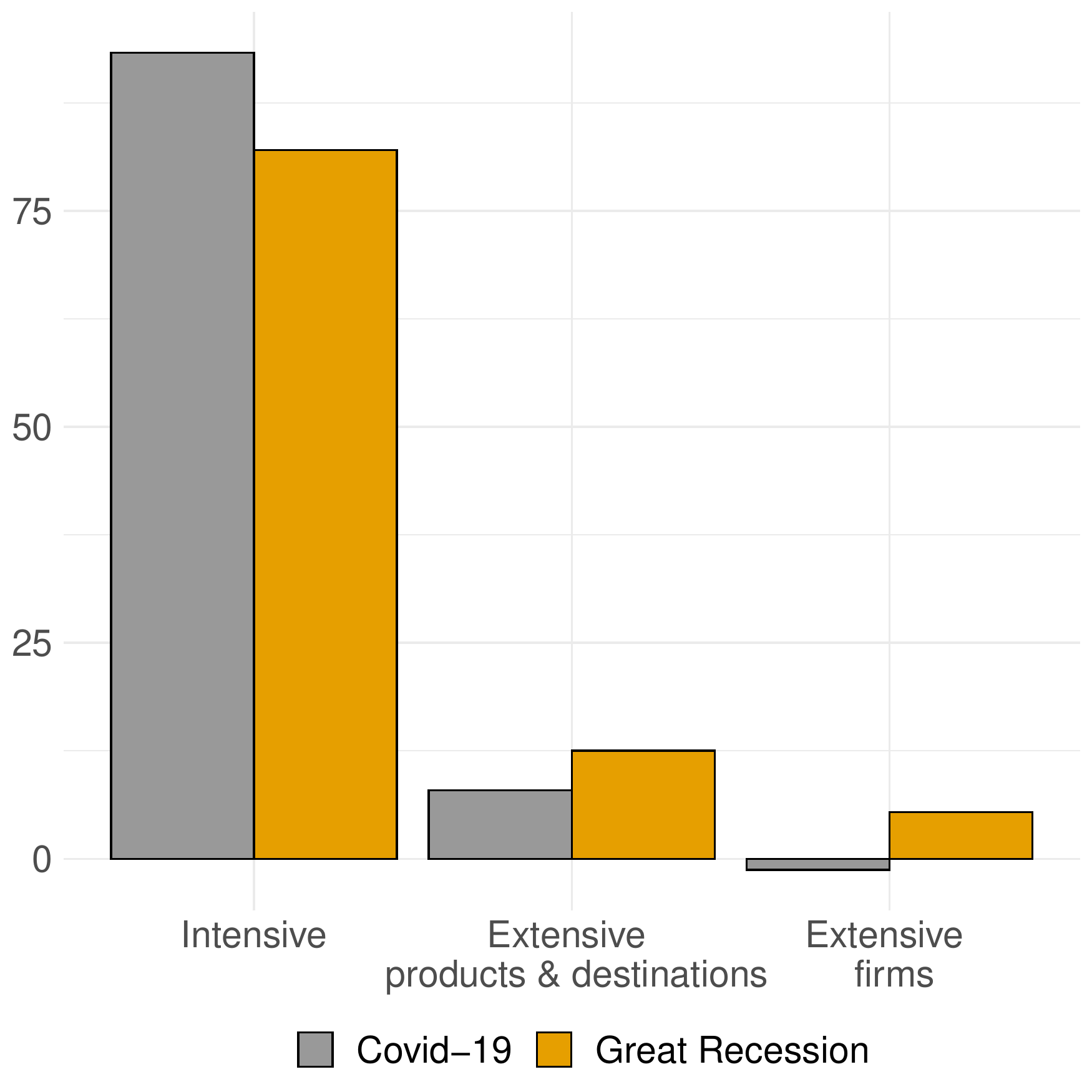}\\
	\end{tabular}
	\end{center}
	\footnotesize{Note: The height of a bar represents the contribution of the margin to the decrease in the value of trade during the crisis. Source: Author's own elaboration based on data from Customs.}
\end{figure}

Extending the analysis of \cite{delucio2020covidcomparacion}, I compute the contribution of the intensive and extensive margins to the decrease in the export and import of goods during the COVID-19 crisis and compare it with that of the Great Recession.\footnote{\label{foot:decomposition}I define products at the Harmonized System 2-digit level. For the Covid-19 crisis, I decompose the change in trade between March-August 2020 and the same period of 2019. For the Great Recession, I compare the period of March-August 2009 with the same period of 2008. See \cite{delucio2020covidcomparacion} for a detailed explanation on how calculations are performed.}  The intensive margin explained 95.3\% of the decrease in exports during the COVID-19 crisis (Figure~\ref{fig:intensive_extensive}). The disappearance of export flows (i.e., combinations of products and destinations) by firms that continued exporting contributed 3.8\% to the decrease. The disappearance of some exporters only accounted for 0.9\% of the decrease. The intensive margin was also the main contributor to the drop in exports during the Great Recession: 95.7\%.\footnote{This percentage differs from the one reported by \cite{delucio2011extensive}, because it only covers six months of the Great Recession, whereas the latter considers the entire crisis.} The disappearance of some export flows by ongoing exporters accounted for 6.3\% of the decline. The contribution of the firm extensive margin was negative (-2.0\%), indicating that the increase in exports by new exporters was larger than the reduction of exports by firms that ceased to export.\footnote{\cite{almunia2018venting} and \cite{delucio2018recession} showed that the drop in domestic demand during the Great Recession led many firms to begin to export.}

The intensive margin was also the main contributor to the decrease in imports during the COVID-19 crisis: 93.3\%. The reduction in the destination and product portfolio was 7.9\%. Interestingly, new importers smoothed the decrease in imports by 1.2\%. The intensive margin was also the main contributor to the drop in imports during the Great Recession, but with a lower percentage: 82.1\%. The drop in imports by ongoing importers was larger during the Great Recession, 12.5\%. As opposed to the COVID-19 crisis, imports by new importers did not compensate the loss of imports by firms that ceased to import. 

The large share of the intensive margin in the decrease of exports and imports, therefore, leads to the anticipation of a rapid recovery of trade flows when the health crisis ends. A feature that can further support a rapid recovery is a larger drop in exports among top exporters than non-top exporters. I define top exporters as the top 1\% of firms that exported more during the March-August 2019 period. These firms contributed with 61\% of Spanish exports during that period. I find that during the COVID-19 crisis, exports declined by 27.7\% among top exporters and 13.3\% among non-top exporters. Because large firms have more resources to recover trade relationships that are lost during the crisis and to create new ones, and have more financial resources to overcome a slump, their exports can recover faster than those of non-top firms.

\section{Did COVID-19 have a larger negative impact on Spanish exports than on other countries'?}
\label{sec:larger}

\begin{figure}[t!]
	\begin{center}
		\caption{Changes in exports during the COVID-19 crisis. Spain vs. other EU countries (March-August 2020 relative to March-August 2019; \%)}
		\label{fig:x_eu}
		\begin{tabular}{c c}
\multicolumn{2}{c}{Panel (A): Goods and services}	 \\
	\multicolumn{2}{c}{\includegraphics[height=2.5in]{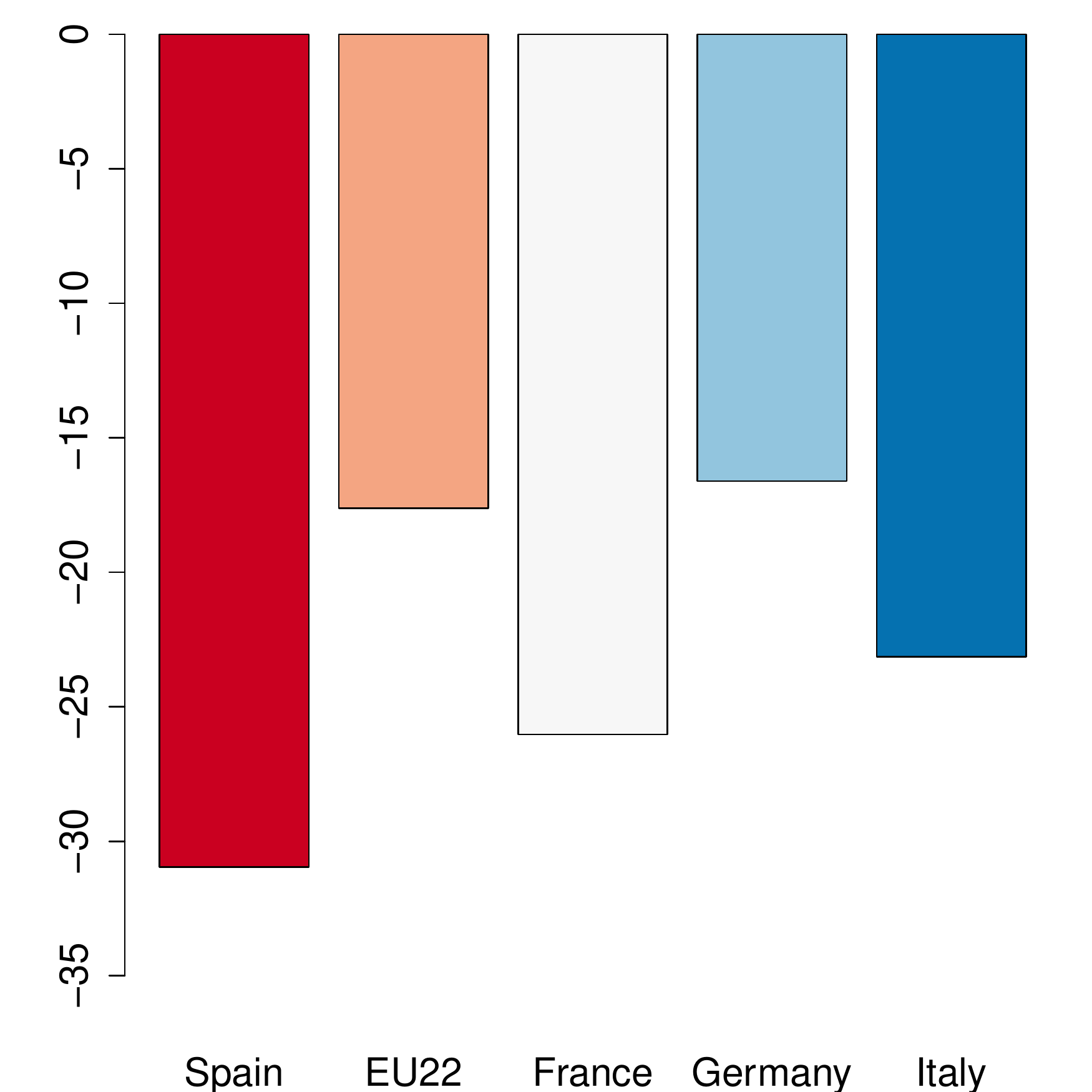}} \\
			Panel (B): Goods & Panel (C): Services\\
		\includegraphics[height=2.5in]{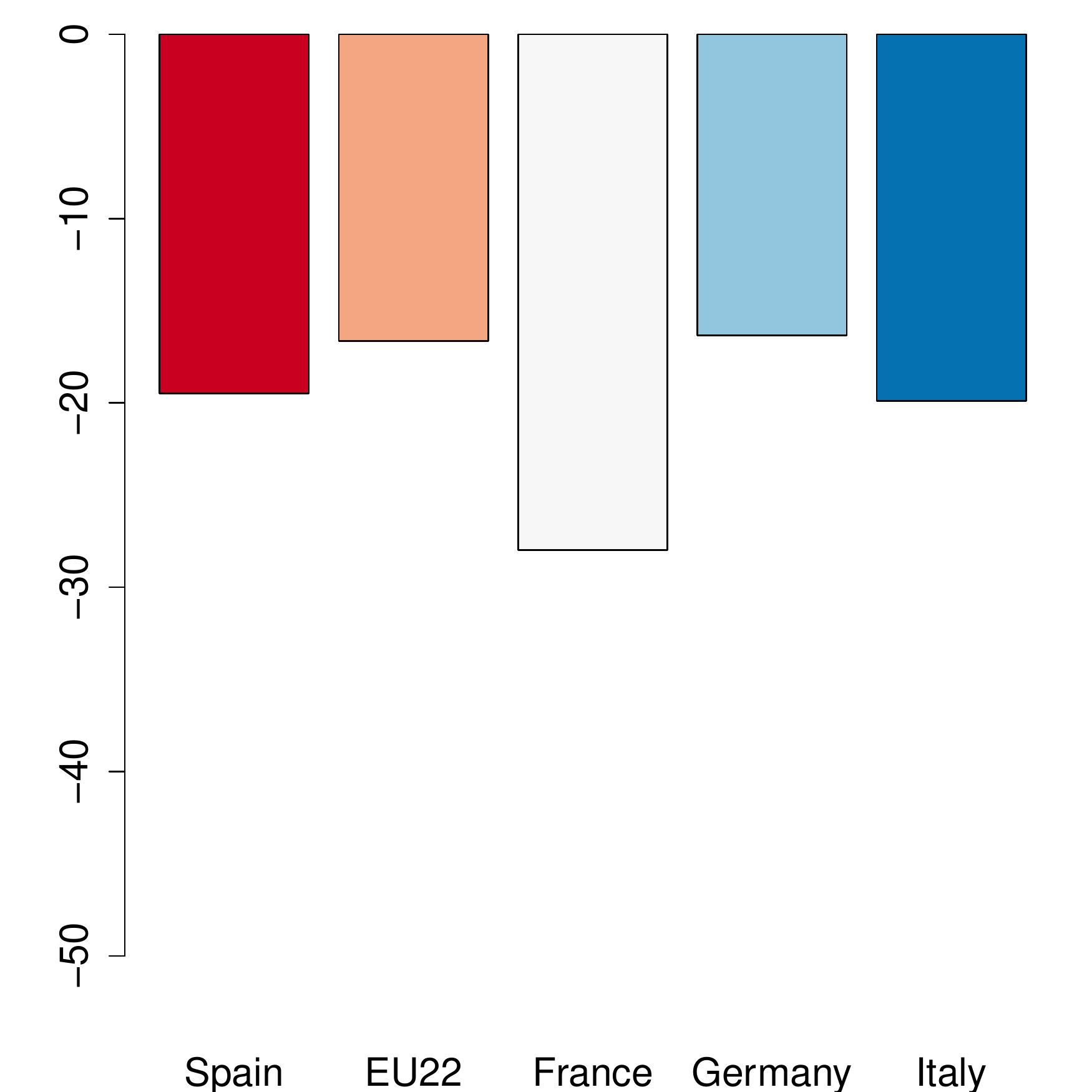} & \includegraphics[height=2.5in]{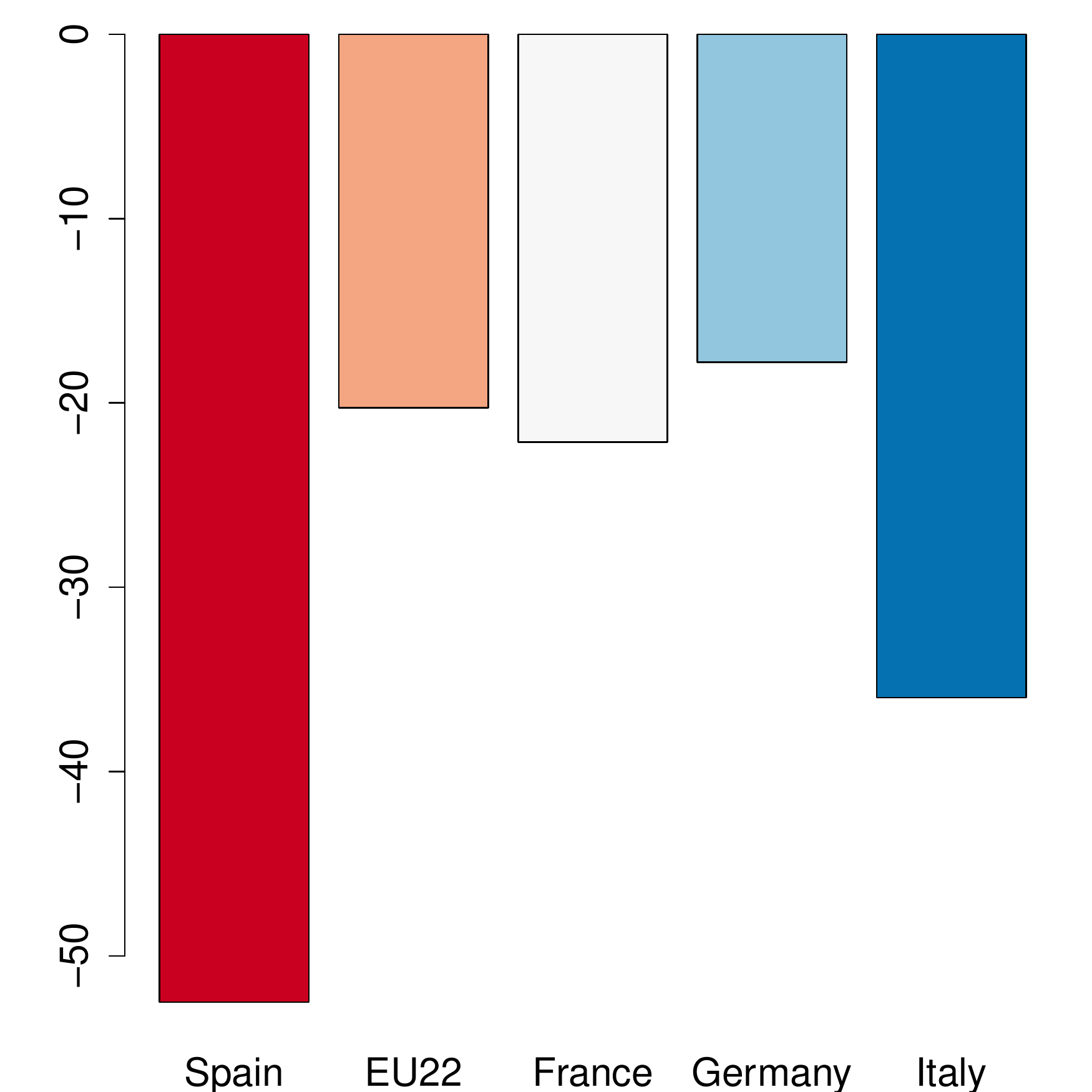} \\	
		\end{tabular}
			\end{center}
	\footnotesize{Source: Author's own elaboration based on data from Eurostat, Spanish Customs, and Bank of Spain.}
\end{figure}

This section explores whether COVID-19 had a larger negative impact on Spanish exports than on those of other countries. Panel~(A) of Figure~\ref{fig:x_eu} shows the decrease in the export of goods and services in Spain, a group of 22 EU (EU22) countries, and the top-3 EU exporters: France, Germany, and Italy.\footnote{The EU22 comprises Belgium, Bulgaria, Czechia, Denmark, Estonia, Finland, France, Germany, Greece, Hungary, Italy, Latvia, Lithuania, Luxembourg, Malta, The Netherlands, Poland, Portugal, Romania, Slovakia, Slovenia, and Sweden.} I compare total exports of March-August 2020 with that of the same period of 2019.\footnote{Monthly data on export of goods and services for EU countries were obtained from Eurostat's Balance of payments by country - monthly data (BPM6) database (\url{https://appsso.eurostat.ec.europa.eu/nui/show.do?dataset=bop_c6_m&lang=en}). Spain's data was not available in this dataset. Therefore, I used the Customs and Bank of Spain's databases mentioned previously.} The decrease in exports in Spain, 31.0\%, was three-quarters larger than that of the EU22 (17.6\%). Exports also fell more in Spain than in Germany (16.6\%), France (26.0\%), and Italy (23.1\%). These figures indicate that COVID-19 had a particularly negative effect on Spanish exports.

Panel~(B) of Figure~\ref{fig:x_eu} shows the decrease in the export of goods. The decrease in Spain (19.5\%) was also larger than that of the EU22 (16.6\%), but the difference was lower than that shown in panel~(A). The export of goods decreased more in Spain than in Germany (16.3\%), was similar to that of Italy (19.9\%), and less than that of France (28.0\%). Panel~(C) shows that the differential impact of COVID-19 on Spanish exports can be explained by the large decrease in service exports (52.5\%), which is more than twice the decrease in the EU22 countries (20.3\%). Italy also experienced a very large reduction in service exports (36.0\%). However, these decreases were much smaller in France (22.1\%) and Germany (17.8\%).

\begin{figure}[htbp]
	\begin{center}
		\caption{Year-to-year change in the export of goods. Top 20 Spanish HS chapters (\%)}
		\label{fig:sitc}
		\includegraphics[width=15cm,height=9cm]{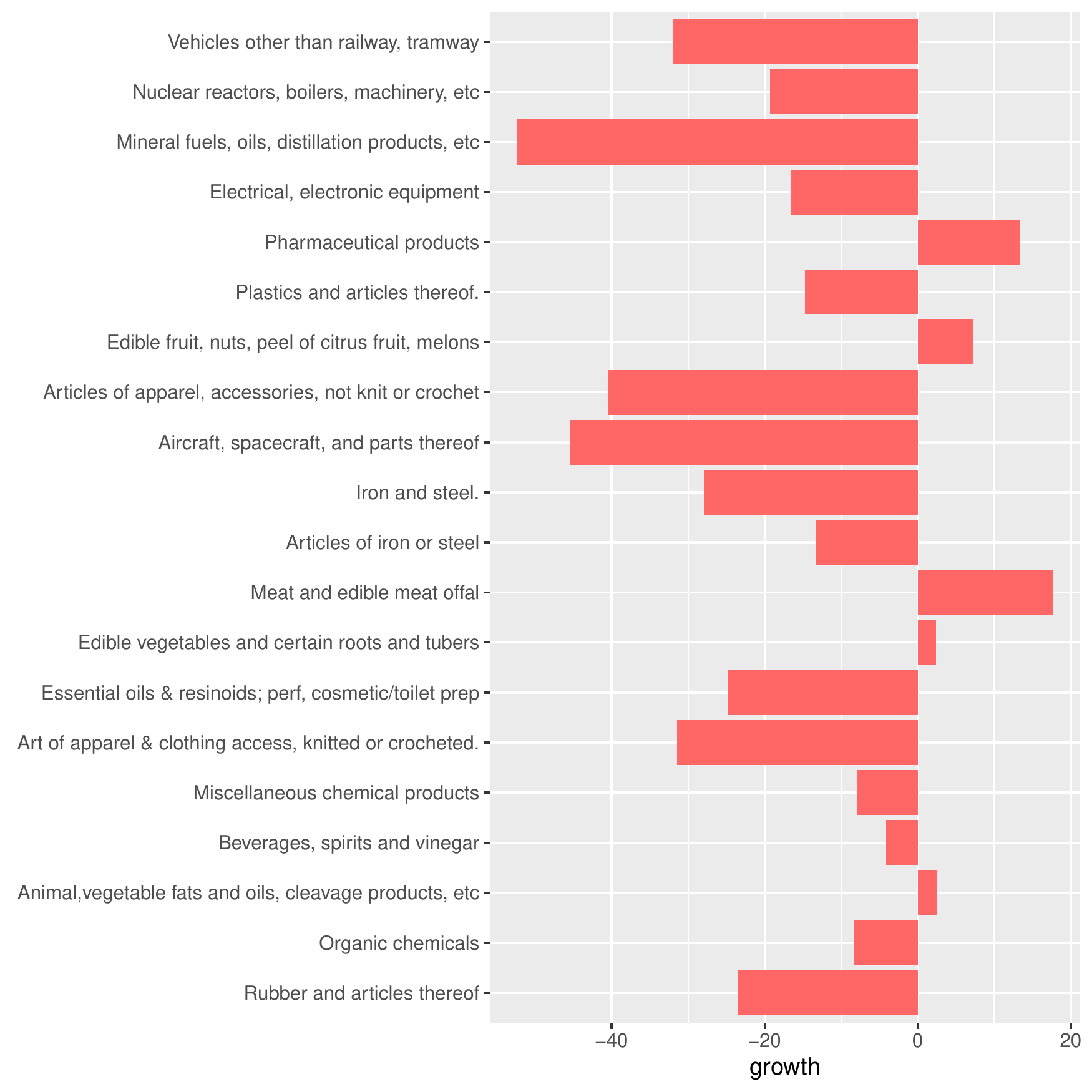}
	\end{center}
	\footnotesize{Source: Author's own elaboration based on data from Customs.}
\end{figure}

To understand the larger decrease in the export of goods in Spain than in EU22, Figure~\ref{fig:sitc} presents the changes in exports from the 20 Harmonized System (HS) chapters
most exported by Spain.\footnote{\label{foot:chapters}I exclude chapter 98 (Complete industrial plants) and 99 (Special combined nomenclature codes) from the analysis, because it is unknown the products that were included in them.} The top-20 HS chapters are ordered from top to bottom, according to their share of the Spanish export of goods in 2019. There was a very large decrease in the export of vehicles, 31.9\%, the most important chapter of Spanish exported goods in 2019 (17.5\%). The drop in the second-most important chapter, machinery, was 19.3\%. There were also very large decreases in fuels (52.3\%), apparel-related chapters (40.5\% and 31.4\%), and aircraft (45.5\%). Contrarily, exports increased in food-related chapters and pharmaceutical products. The specialization in transport equipment, capital goods, and outdoor goods explains the larger negative effect of COVID-19 on Spanish export of goods than on that of EU22 countries.\footnote{\label{foot:outdoor}An outdoor good is a product that is mostly consumed outside the household. For example, ski-boots are outdoor goods, while slippers are not. The list of outdoor goods comes from \cite{delucio2020covidcollapse}. They showed that the export value of goods that are consumed outdoors decreased five times more than that of the remainder consumption goods.}

The larger decrease in the export of services in Spain can be explained by the larger share of tourism in Spanish exports. According to the Organization for Economic Cooperation and Development data, personal travel accounted for 52.0\% of Spanish service exports in 2018, whereas it was only 16.9\% for the EU28. As shown in Section~\ref{sec:chronology}, tourism was the most affected industry by the COVID-19 crisis. 

Finally, I explore whether the country-distribution of exports also contributed to the differential negative impact of COVID-19 on Spanish trade.\footnote{See \cite{alvarez2018diversification} for an analysis on how the change in the destination of Spanish exports contributed to the evolution of Spain's share in the global export of goods.} Owing to the lack of data on services, I only perform this analysis for goods. I analyze what would have been the decrease in the export of goods if the distribution of Spanish foreign sales by destinations was the same as that of Germany, France, and Italy. I selected these countries because they are major exporters, are located in Europe, belong to the EU, and share the same currency. Thus, the gravitational forces affecting the destinations of the exports of these reference countries were similar to those influencing the portfolio of Spanish export partners. I group destinations into the following seven areas: EU + United Kingdom, Rest of Europe, Africa, Asia, North America, Rest of America, and Oceania. I find that the decrease in Spanish export of goods would have been one percentage points lower if the distribution of exports by destination had been the same as that of France, Germany, and Italy.\footnote{\label{foot:destinations} I add-up the value of exports of the three countries in 2018 and 2019 by zone. Then I calculate the share of each zone in total exports.} The larger negative impact was caused  by the higher concentration of Spanish exports than that of the reference countries in EU+UK markets (66.3\% vs. 53.5\%), Africa (6.8\% vs. 2.7\%), and the Rest of America (3.7\% vs 1.5\%), and the lower growth of exports in some of these zones than in the remainders.\footnote{Exports to Africa decreased year-to-year by 26.1\%, to Asia by 15.8\%, to the EU+UK by 17.3\%, to the rest of Europe by 12.2\%, to North America by 22.2\%, to the rest of America by 32.1\%, and to Oceania by 26.9\%.}

\section{Impact of COVID-19 by region}

\begin{figure}[htbp]
	\begin{center}
		\caption{Impact of COVID-19 on provinces exports (\% change)}
		\label{fig:provinces}
		\vspace{-1.5cm}
		\includegraphics[height=5.5in]{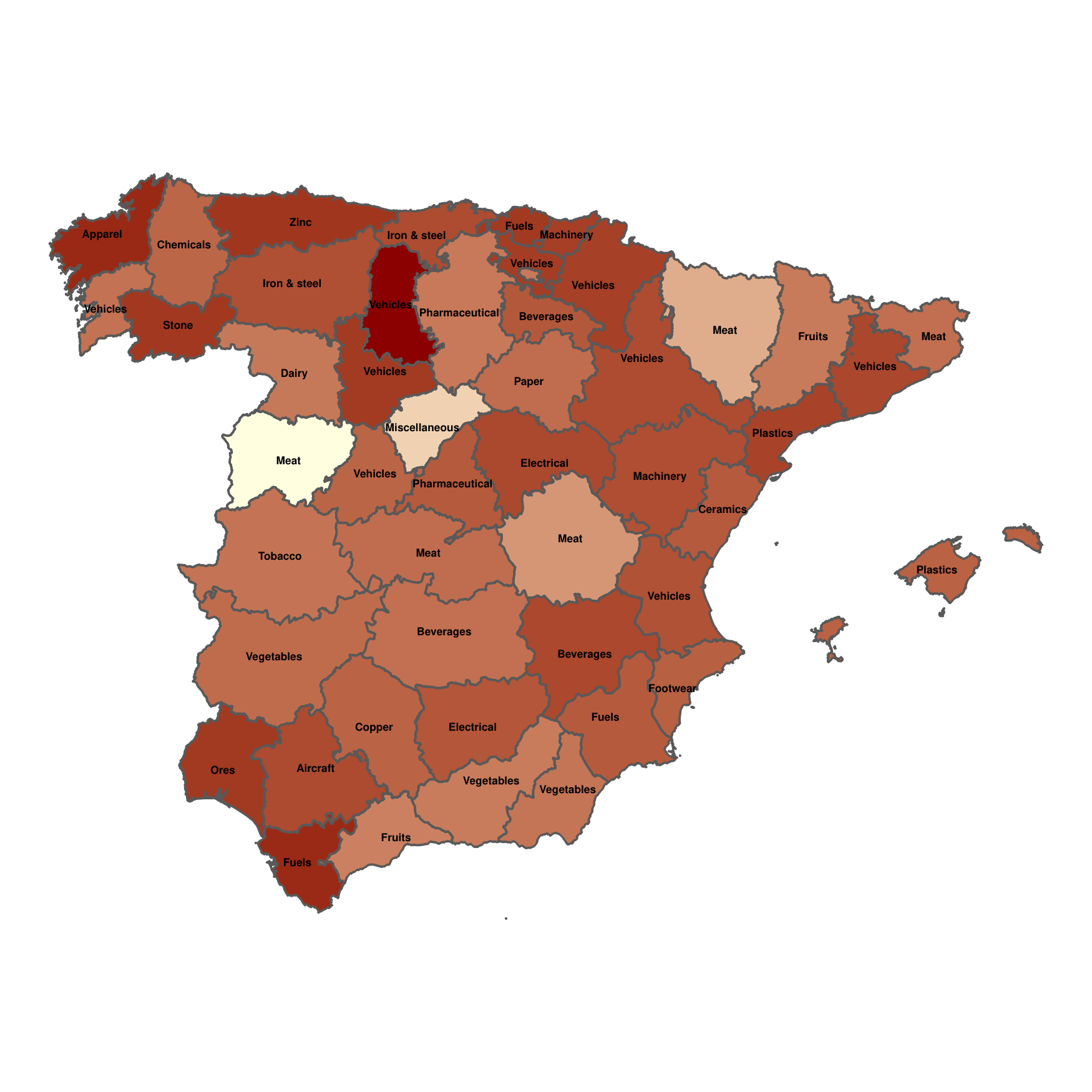}
	\end{center}
\vspace{-2cm}
	\footnotesize{Note: the word written in the center of each province is the HS chapter having the highest share of the province's exports in 2019. Source: Author's own elaboration based on data from Customs.}
\end{figure}

The impact of the COVID-19 crisis varied across Spanish provinces according to their export specializations.\footnote{For an analysis of the impact of COVID-19 on the GDP of Spanish regions, see \cite{prades2020covid}.} Figure~\ref{fig:provinces} shows a heat map of the impact of COVID-19 on the export of goods by Spanish provinces. Larger decreases in exports are represented by darker colors and smaller decreases by lighter colors.\footnote{I do not include the Canary Islands or the autonomous cities of Ceuta and Melilla.} For each province, I also annotate the HS chapter having the highest share of exports in 2019. 

Darker colors represent provinces that specialized in vehicles, fuels, and apparel. Contrarily, provinces with lighter colors specialized in foodstuff, pharmaceuticals, and miscellaneous manufacturing.\footnote{The largest growth in exports occurred in the province of Salamanca. The product having the highest share of exports in that province in 2019 was meat. However, the growth in exports in 2020 can be explained by nuclear fuel.} These patterns are in line with Figure~\ref{fig:sitc}, which shows the HS chapter impact of COVID-19. Although there are no data on the regional distribution of service exports, it is reasonable to expect that the most affected regions were the Balearic and Canary Islands, where tourism represented 35\% and 45\% of GDP in 2018, respectively.\footnote{Data from Exceltur. Available at: \url{https://www.exceltur.org/pib-y-empleo-turistico-por-c-c-a-a/}}

\section{Import dependence and trade in medical material and equipment}
\label{sec:global_value_chains}

As mentioned in the introductory section, the COVID-19 crisis manifested, economically, as a supply crisis in Spain, highlighting the dependence of some Spanish firms on just a few importers. \cite{delucio2020postalde} calculated the number of supplier countries per product among Spanish importers of goods in 2019. They identified products using an eight-digit code in the Combined Nomenclature. They found that the median Spanish importer sourced a product from only one country. The most repeated single-source country was China (26\%), followed by Germany (12\%), and Italy (11\%). If imports by firms that only sourced from a single country were totaled, they would represent 19\% of all Spanish import of goods in 2019.  

This begs the question of whether this import dependency made Spain more vulnerable to the COVID-19-induced value-chain disruption. Although at the beginning of the pandemic some firms were affected by a lack of material coming from China and northern Italy, a survey of regular exporters concluded that the drop in foreign demand was the most important factor explaining the drop in exports during the second quarter of 2020 \citep{industria2020coyuntura2t}.\footnote{In another survey, firms considered that a decrease in foreign demand would have a larger impact on their activity than would disruptions in the supply chain \citep{ine2020confianza}. \cite{imf2020weoctober} also contended that the decline in trade volumes largely reflected weak final demand from customers and firms.} \cite{delucio2020covidcollapse} provided econometric evidence that the containment measures adopted by countries where Spanish imports originated had no effect on the value imported by Spanish firms. However, they increased the probability of a firm terminating their import relationship. \cite{bonadio2020covid} estimated the contribution of global supply chains to the COVID-19-induced reduction in GDP in 64 countries. The authors considered the percentage of tasks that could be performed online, the stringency of confinement measures, and domestic firms' dependency on foreign suppliers, and estimated that COVID-19 caused a 34.6\% contraction of GDP in Spain, of which 9.8 percentage points (28\%) could be explained by the disruption of global value chains. They also calculated the potential contraction of GDP had Spain produced all intermediate inputs at home. They found that the contraction would have been larger: 36.3\%. This can be explained by the more stringent confinement measures adopted by Spain than those by countries supplying intermediate inputs to Spain. 

Concerns about the risks of relying on foreign suppliers were especially intense at the beginning of the COVID-19 crisis because they were related to the sourcing of medical materials and equipment (e.g., protective gear and medical ventilators), which were key fighting the pandemic. For example, to ensure that domestic production was dedicated to local needs, in March 2020, the EU introduced a regulation making exports of certain products subject to authorization.\footnote{The EU 2020/402 regulation came after France and Germany introduced barriers to the export of medical equipment that also affected other EU countries. In exchange for removing those barriers, the EU imposed regulations on exports to non-EU countries \citep{felbermayr2020covid}. This regulation was applied for a six-weeks period, and was further extended for a period of 30 days by regulation 2020/568 (April 23rd, 2020). It ceased to apply on May 26th, 2020.} In the scramble to acquire needed medical equipment, some EU members temporarily seized shipments destined to other member states, were outbid by contenders when shipments were bound to take off from source countries, or purchased products that were defective or below standard \citep{elpais2020guerramascarillas,bbcnews2020countriesrejectChina}. 

\begin{table}[t]
\centering
\caption{Import dependence on medical material and equipment} 
\label{tab:medical}
\begin{tabular}{lrrllr}
  \toprule
  Category & Imports/ & Import& Main & Main single& Share \\
 &production&growth&supplier&supplier&in total\\
 \midrule
Protective garments & 10.0 & 430.7 & China & China & 53.8 \\ 
  Oxygen therapy & 12.9 & 59.2 & China & China & 21.9 \\ 
  Test kits & 5.4 & 27.8 & Germany & China & 22.7 \\ 
  Disinfectants & 1.3 & 11.7 & United States & France & 10.7 \\ 
  Other medical devices & 8.8 & 0.7 & Germany & China & 20.7 \\ 
  Medical consumables & 2.3 & -8.5 & Germany & China & 21.0 \\ 
   \bottomrule
\end{tabular}
\caption*{\begin{footnotesize}Note: The ratio of imports to domestic production (imports/production) represents 2018. Import growth calculates changes in imports between March and August 2020 and the same period of 2019. The main supplier is the largest supplier in the category during the March-August 2020 period. The main single supplier was the country most repeated among importers having a single supplying country per product as of March-August. Share in total was calculated using imports by firms having only one supplier over total imports. Source: author's own elaboration using data from Prodcom and Customs.\end{footnotesize}}
\end{table}


Spain was especially vulnerable to the COVID-19 crisis because the domestic capacity could not respond to the increase in demand for medical materials and equipment. Table~\ref{tab:medical} shows the ratio of imports to domestic production (imports/production) before the COVID-19 crisis for the six categories identified as critical to combating the COVID-19 pandemic by the World Customs Organization (WCO) and the World Health Organization (WHO).\footnote{Production data come from the Eurostat Prodcom Database and correspond to 2018. The products included in each critical category and their HS six-digit codes were obtained from \cite{wco2020covidmedical}. I linked the HS codes with Prodcom using a correspondence table from the Reference and Management of Nomenclatures (\url{https://ec.europa.eu/eurostat/ramon/relations/index.cfm?TargetUrl=LST_REL}).} I sorted the categories, from top to bottom, by the growth in imports during the COVID-19 crisis, relative to the same period of 2019. The growth in imports/production ratio was very high for protective garments, oxygen-therapy equipment, and tests kits (the three categories that experienced the largest growth of imports during the pandemic). The growth in protective garment imports was impressive and stark: 430.7\%. Nevertheless, most of this increase can be explained by the surge in the price of imports. The kilograms of imported garments only rose by 93\%. The dependence ratio was also large for other medical devices. However, imports increased very little in this category during the pandemic. Germany was the main supplier of imports during the COVID-19 crisis in the following three categories: test kits, other medical devices, and medical consumables. China was the main supplier of protective garments and oxygen therapy. The United States was the top supplier of disinfectants. However, if we analyze the most important source for importers that only relied on a single supplier, China was the most repeated country. The value of imports that were sourced by importers with a single source country per product over total imports was very high for protective garments and non-negligible for the remaining categories.

\section{Will COVID-19 change Spanish trade?}
\label{sec:change}

Richard Baldwin argued that \enquote{things are the way they are for good reasons. If the reasons don’t change permanently, the outcomes won’t change permanently} \citep{baldwin2020covid}. In this section, I speculate whether COVID-19 has altered some of those ``goods reasons'' and if it will lead to permanent changes in Spanish trade. 

COVID-19's clearest impact was the reduction of people's mobility. Firms responded to this by moving many activities online. Domestic equipment and internet connections were improved, and workers quickly learned how to be productive remotely. If firms can learn how to manage remotely-provided day-to-day activities, they might decide to further outsource those jobs to personnel located in lower-wage countries \citep{baldwin2019globotics}. In parallel, the digital transformation accelerated by COVID-19 might cause firms to be more open to hiring business services from foreign countries. This could increase Spanish exports in services, such as market research or engineering, which can be provided online and where the share of exporters is large \citep{minondo2015servicios}.

Contrarily, services requiring the movement of people have been very negatively affected by COVID-19. Until a vaccine is distributed, restrictions to mobility will remain. This and people's ongoing concerns about health conditions in destination countries will continue to discourage tourism and similar provision of services that require the movement of people. In any case, a vaccine notwithstanding, COVID-19 has shown that tourism is very fragile to pandemics. Firms will explore business models characterized by a lower dependence on long-range foreign tourists and overcrowding. This transition may be particularly difficult for Spain, where tourism represented 12.3\% of its GDP and foreign tourists accounted for 75\% of the total expenditure in this industry.\footnote{INE, Cuenta Sat\'{e}lite del Turismo en Espa\~{n}a, Encuesta de Turismo de Residentes, and Encuesta de Gasto Tur\'{i}stico. GDP data are from 2018; share of foreign tourists in total expenditure data are from 2019.} 

Restrictions to the movement of people could also have a negative impact on trade in the medium and long runs in Spain. \cite{delucio2011extensive} showed that as new and incumbent exporters' diversified into new destinations and products, they contributed by half to the growth of Spanish exports between 1997 and 2007. To grow along these extensive margins, Spanish firms must build trust-based relationships with new foreign customers. These relationships rely on a social rapprochement that cannot be successfully substituted with video calls yet.\footnote{Despite the improvement in communication brought by the internet, business trips continued to rise \citep{cristea2011tradetravel,coscia2020travel}.} If COVID-19 continues to increase the costs of traveling, international business meetings and fairs will continue to diminish, reducing Spanish firms' opportunities for opening new trade relationships.

It is also likely that countries will rely less on foreign suppliers to source essential products. To address future crises, countries can build strategic inventories of essential equipment, while keeping production in countries that have a comparative advantage in these products.\footnote{Switzerland holds reserves of food and medicines to cope with unseen crises. During the COVID-19 crisis, confidence on the availability of those reserves precluded the panic purchases observed in other countries \citep{ft2020switzerland}.} However, future health crises may demand new types of medical devices, medicines, or tests that can only be developed if local capacities are present. This dynamic perspective may lead countries to trade efficiency for domestic reaction capacity, nurturing domestic skills through contracts between the public health sector and local firms.\footnote{See \cite{guinea2020dependency} for a critical view on this strategy.} For example, on August 6th, 2020, the US President issued an executive order to ensure that essential medicines would be produced in the US.\footnote{\url{https://www.whitehouse.gov/presidential-actions/executive-order-ensuring-essential-medicines-medical-countermeasures-critical-inputs-made-united-states/}} Such nurturing of domestic capacities in health-related products is consistent with previous policies aiming to develop strategic sectors (e.g., electric cars, digital-intensive goods and services, and green technologies). For Spain, it is likely that the drive for higher domestic capacity will be framed under EU's open strategic autonomy framework, which could lead to an increase in the share of this economic area of Spanish trade \citep{eucommission2020industrialstrategy}.

The regionalization of Spanish trade might also be favored by the changes that COVID-19 might cause in global value chains. The Tohoku earthquake and tsunami in Japan in 2011 and the floods in Thailand during the same year already revealed many of the vulnerabilities of global value chains to natural disasters \citep{boehm2019tohokuearthquake,haraguchi2015flood}. However, the global scope of COVID-19 rendered the potential risk of disruption much more apparent to firms.\footnote{\cite{kilic2020covid19automatization} argued that firms could respond to potential disruptions in the supply chain by substituting workers with COVID-19-free machines.} 

\cite{diazmora2020globalvaluechains} argued that firms could trade off cost efficiencies for supply chain shock resilience by keeping larger inventories, investing in technologies that closely monitor supplier situation, and diversifying their pool of suppliers. Furthermore, they argued that firms might want to locate their supply chains geographically closer. First, difference of labor costs between China and developed countries has decreased, and increased automation could make these differences even less relevant. Second, geographically-close suppliers would allow firms to react quickly to changes in demand. Third, policies to avert climate change could eventually lead firms to save on transportation costs. Finally, firms will likely choose suppliers belonging to the same regional trade agreements, because they are less likely to change their trade policies unilaterally. 

\cite{gandoy2020nota} argued that Spain could benefit from this supply chain regionalization process because of its competitive labor costs, relative to other EU countries, good infrastructures, and labor skills. However, the regionalization of supply chains might harm Spanish firms that supply industrial equipment to customers outside Europe. Potential customers might doubt that Spanish firms will be able to supply their goods owing to pandemic-related disruptions. This concern will be especially acute for large-maturation projects, which require the movement of personnel to fine-tune equipment at the customer's premises. 

In any case, finding new suppliers is not an easy task. Firms invest heavily in identifying suppliers that will meet their specifications, quality standards, and shipment schedule and will have the capacity to respond rapidly to changes in production planning. Suppliers often must make specific investments to adapt their product to customers' needs. Furthermore, there are many elements in the supplier-customer relationships that cannot be written into contracts or enforced outright. Thus, firms must be confident that other parties will not take advantages. All these reasons lead to sticky supplier-customer relationships \citep{worldbank2020valuechains}. Moreover, the swift recovery in the trade of goods anticipated in Section~\ref{sec:versus} of this document relies on the stickiness of the customer-supplier relationships.

\section{Conclusion}
COVID-19 has caused the steepest decrease in Spanish trade in decades. Trade fell for both goods and services. However, the containment measures adopted to arrest the spread of the virus made the decrease especially intense for services. Spain's export specialization in products that facilitate the movement of people and goods, capital goods, outdoor products, and services that rely on the movement of people has made the negative impact of COVID-19 on trade greater than in other large EU countries. The nature of the trade collapse in goods, which concentrated on the intensive margin and particularly affected top exporters, suggests a rapid recovery of trade flows when health risks disappear. However, the recovery of tourism might take longer because it depends on the perceptions of foreign customers and their governments on the health situation in Spain.

In addition to the short-term massive reduction in trade flows, COVID-19 might also have longer-term impacts. Spain must develop strategies to reduce its dependence on foreign tourism receipts and exploit opportunities to expand business service exports. In the medium and long terms, Spanish trade might grow less owing to the higher costs of establishing new trade relationships. Furthermore, the development of domestic capacity in strategic sectors and the regionalization of value chains could lead to further increases in the shares of EU countries hold in Spanish trade. 

Finally, the difficulties of obtaining medical materials and equipment to fight COVID-19 have triggered a general mistrust in the capacity of trade to adequately respond to countries' needs in the face of emergencies. This concern will exacerbate if countries do not manage the future distribution of the COVID-19 vaccine wisely. Failures in that case would quickly reinforce isolationist positions that would surely damage future trading relationships.

\end{document}